\documentclass[a4paper,12pt]{article}
\usepackage[esperanto,portuguese,english]{babel}
\usepackage{epsfig}
\usepackage{icomma}
\usepackage{float}
\usepackage[utf8]{inputenc}
\usepackage[T1]{fontenc}
\usepackage{lmodern}

\newcommand{\ppparallel}[1]{} \newcommand{\ppmaldekstra}[1]{} \newcommand{\ppdekstra}[1]{}
\usepackage{parallel} \renewcommand{\ppparallel}[1]{#1} % bilingue
\newcommand{\comentario}[1]{}

\newcommand{\dd}{\mathrm{d}}
\newcommand{\jj}{\^{\j}} 

\ppparallel{
\newlength{\pplw}\setlength{\pplw}{0.47\textwidth}%0.453
\newlength{\pprw}\setlength{\pprw}{0.49\textwidth}%0.507

\newcommand{\ppn}{\noindent}              %p/ nao fazer parag
\newcommand{\ppl}[1]{\ParallelLText{\selectlanguage{esperanto}#1}}
\newcommand{\ppr}[1]{\ParallelRText{\selectlanguage{english}#1}\ppp}
\newcommand{\ppln}[1]
{\ParallelLText{\ppn \selectlanguage{esperanto}#1}} %p/ nao parag
\newcommand{\pprn}[1]
{\ParallelRText{\ppn \selectlanguage{english}#1}\ppp} %p/ nao parag

\newcommand{\bea}{\vspace{-1ex}\begin{eqnarray}}
\newcommand{\eea}{\end{eqnarray}}
}

\ppmaldekstra{
\newcommand{\ppl}[1]{\selectlanguage{esperanto}#1}
\newcommand{\ppln}[1]{\noindent \selectlanguage{esperanto}#1}
\newcommand{\ppr}[1]{\selectlanguage{english}}
\newcommand{\pprn}[1]{\noindent \selectlanguage{english}}

\newcommand{\bea}{\begin{eqnarray}}
\newcommand{\eea}{\end{eqnarray}}
}

\ppdekstra{
\newcommand{\ppl}[1]{\selectlanguage{esperanto}}
\newcommand{\ppln}[1]{\noindent \selectlanguage{esperanto}}
\newcommand{\ppr}[1]{\selectlanguage{english}#1}
\newcommand{\pprn}[1]{\noindent \selectlanguage{english}#1}

\newcommand{\bea}{\begin{eqnarray}}
\newcommand{\eea}{\end{eqnarray}}
}

\title{{Tempa voja\^go kaj geodezioj en \^generala relativeco \ppparallel{\\Time travel and geodesics in general relativity}}}
\author{F.M. Paiva \\ 
{\small Departamento de F\'isica, Unidade Humait\'a II, Col\'egio Pedro II} \\
{\small Rua Humait\'a 80, 22261-040  Rio de Janeiro-RJ, Brasil; fmpaiva@cbpf.br} 
\vspace{.7ex} \\
A.F.F. Teixeira \\
{\small Centro Brasileiro de Pesquisas F\'isicas} \\
{\small 22290-180 Rio de Janeiro-RJ, Brasil; teixeira@cbpf.br}}
\begin{document}
\selectlanguage{esperanto}
\maketitle
\thispagestyle{empty}

\begin{abstract}\selectlanguage{esperanto}
\^Ce la \^Generala Relativeco, en homogena metriko de Som-Raychaudhuri, ni studas geodeziojn de la tri tipoj: tempa, nula, kaj spaca, speciale la malmulte konatajn samtempajn geodeziojn. Ni anka\u u studas ne-geodezian cirklan movadon kun konstanta rapido, speciale fermitajn kurbojn de tempa tipo, kaj movadon de voja\^ganto al estinto. 
  
\ppparallel{\selectlanguage{english}
In the homogeneous metric of Som-Raychaudhuri, in general relativity, we study the three types of geodesics: timelike, null, and spacelike; in particular, the little known geodesics of simultaneities. We also study the non-geodetic circular motion with constant velocity, particularly closed timelike curves, and time travel of a voyager.} 

\end{abstract}

\ppparallel{
\begin{Parallel}[v]{\pplw}{\pprw}
%\begin{Parallel}[v]{}{}
}

\ppparallel{\section*{\vspace{-2em}}\vspace{-2ex}}   %PORQUE PRECISO DISTO ?

\vspace{5mm}
\ppln{{\bf {\Large 1 \hspace{5mm}Enkonduko}}}
\pprn{{\bf {\Large 1 \hspace{5mm}Introduction}}}
\ppln{} \pprn{}
%\ppsection[0.6ex]{Enkonduko\label{Enkonduko}}{Introdu\c c\~ao}                           Sekcio 1
\ppl{\^Cu estas ebla, ke voja\^ganto revenas al deirpunkto en anta\u ua momento? Teknike dirante:  \^cu kauza malobeo estas ebla? {\mbox Newtona} mekaniko respondas {\it ne}, sed \^Gene\-rala Relativeco diras {\it jes}. \^Sajnas ke G\"odel~\cite{Godel} unue priskribis sistemon kun tio ebleco. Aliaj fizikaj sistemoj, nomitaj modeloj de G\"odela tipo, estis poste  studitaj. En ili, la materio povas reveni al estinto, sed havante kelkan ne-gravitan akcelon. Do \^gia movado estus {\em ne-geodezia}.}
\ppr{Is it possible that a voyager comes back to the starting point in a moment prior to his departure? In technical terms: is causality violation possible? Newtonian mechanics answers {\it no}, but general relativity says {\it yes}. It seems that G\"odel~\cite{Godel} first described a system with that possibility. Other physical systems, called G\"odel-type models, were later studied. In these studies, matter can travel to the past, but having some non-gravitational acceleration. So its motion would be {\em non-geodetic}.}

\ppl{\^Ci tiu artikolo studas geodeziajn kaj ne-geodeziajn movadojn en speciala modelo de G\"odela tipo: la universo de Som-Raychaudhuri~\cite{SomRaychaudhuri}. Ni priskribas la geodeziojn de la tri tipoj. Grava speciala okazo de spaca tipo estas priskribita: la samtempajn geodeziojn, konsistanta el samtempajn najbarajn eventojn, nure. Ni plue montras ke ^ciu geodezia movado de materio a\u u lumo obeas kauzecon. Speciala klaso de ne-geodezia movado estas poste studita, prezentante movadojn kun reveno al estinto.}
\ppr{This article studies geodetic and non-geodetic motions in a particular G\"odel-type model: the universe of Som-Raychaudhuri~\cite{SomRaychaudhuri}. We describe the three types of geodesics: timelike, null, and spacelike. An important special case of spacelike geodesic is described: the geodesics of simultaneities, consisting on simultaneous neighbor events, only. We further show that geodetic motions of matter or light satisfy causality. Then we study a special class of non-geodetic motions, and present motions of matter with time travel.}

\ppl{La linielemento de Som-Raychaudhuri estas skribita en la formo~\cite{SomRaychaudhuri}}
\ppr{The line element of Som-Raychaudhuri is written in the form~\cite{SomRaychaudhuri}}

\bea                                                                                 \label{SR1}%01
\epsilon(\dd s)^2=[\,c\dd t-(\Omega r^2/c)\dd\varphi\,]^2-(\dd r)^2-r^2(\dd\varphi)^2-(\dd z)^2\,, 
\eea

\ppln{kie $\epsilon=+1, 0, -1$ por intervaloj de tempa tipo, a\u u nula, a\u u spaca, respektive. La\u u la \^generala relativeco, la materio rilata al tiu metriko estas elektrizita polvo, kun unuformaj masdenso kaj \^sargdenso. La polvo restas relative al la spaca teksa\jj o, kaj la Lorentza forto en iu polvero estas nula.}
\pprn{where $\epsilon=+1, 0, -1$ for timelike, or null, or spacelike intervals, respectively. According to general relativity, the matter related to this metric is electrically charged dust, with uniform densities of matter and charge. The dust is at rest in the spatial frame, and the Lorentz force upon each dust grain is null.} 

\ppl{\^Car $g_{00}=1$\,, la koordinathorlo\^goj estas normhorlo\^goj. La kinematika parametro de rotacio valoras $\Omega$\,, kaj estas spactempe konstanta. Do, najbare al origino, la spaca teksa\jj o $[r, \varphi, z]$ rotacias kun angula rapido $\Omega=$ konst \^cirka\u u akso $z$\,, relative al inercia kompaso. Ni supozas $\Omega\!>\!0$, implicante rotacion de teksa\jj o en malhora direkto, kiel figuro~\ref{Omega} montras. Tamen, tiu elekto ne implicas fizikan limigon.}
\ppr{Since $g_{00}=1$\,, the coordinate clocks are standard clocks. The kinematic parameter of rotation is $\Omega$\,, and is constant in all spacetime. So, in the vicinity of the origin, the spatial frame $[r, \varphi, z]$ rotates with angular velocity $\Omega=$ const around the $z$-axis, relative to a compass of inertia. We assume $\Omega\!>\!0$, which implies rotation of the frame in anticlockwise direction, as figure~\ref{Omega} shows. However, that choice does not imply loss of physical generality.} 

\begin{figure}[ht]                                                                        %Figura 1
\centerline{\epsfig{file=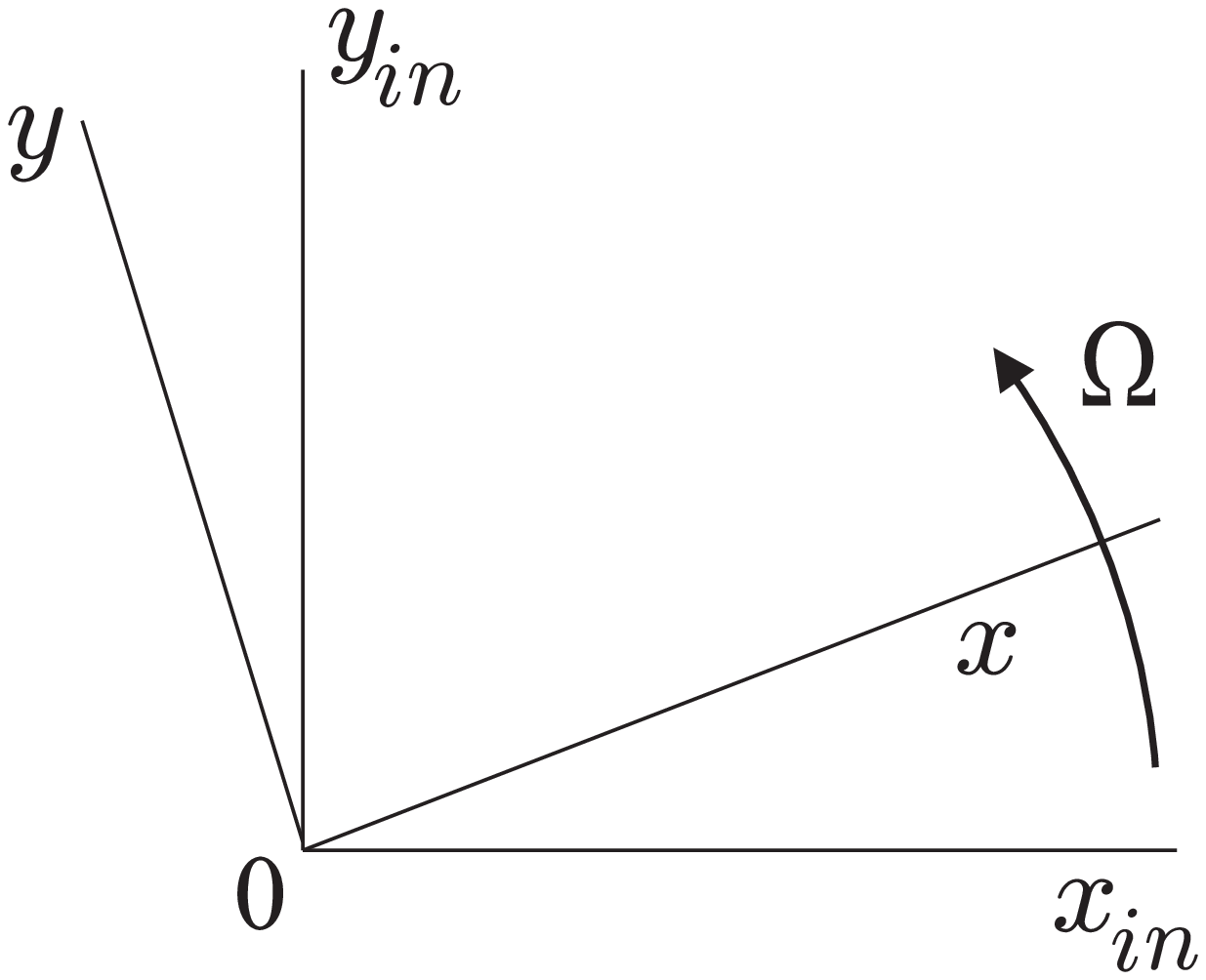,width=4cm}} 
\caption{\^Cirka\u u la origino, la spaca teksa\jj o de Som-Raychaudhuri [$x, y$] rotacias kun angula rapido $\Omega$ relative al spaca teksa\jj o [$x_{in}, y_{in}$], en malhora direkto.
\newline 
Figura~\ref{Omega}: Near the origin, the spatial frame of Som-Raychaudhuri [$x, y$] rotates with angular velocity $\Omega$ relative to an inertial frame [$x_{in}, y_{in}$], in the anticlockwise direction.}
\label{Omega} 
\end{figure}

\ppln{$\;\;\;$ En la sekvanta sekcio ni komencas solvi geodeziajn ekvaciojn. En sekcio~3 ni studas geodeziojn (helicojn) de la tri tipoj, \^cirka\u u la akso $z$, kaj en~4 la samtempajn geodeziojn (rektajn). En sekcio~5 ni studas cirklajn movadojn de voja\^ganto al sia pasinto. Kaj en~6 ni konkludas kaj prezentas interesajn faktojn. Sekve ni difinas kelkajn utilajn kvantojn, detalotaj en~\cite{reltemp2}.}
\pprn{$\;\;\;$ In the next section we start solving the geodetic equations. In section~3 we study  geodesics (helices) of the three types, around the $z$-axis, and in section~4 the (straight) geodesics of simultaneities. In section~5 we study circular motions of a voyager towards his past. And in 6 we conclude and present interesting facts. In the following we define some useful quantities, to be detailed in~\cite{reltemp2}.} 

\ppln{$\;\;\;$ {\it Distanco} inter du najbaraj punktoj en spaca teksa\jj o:}
\pprn{$\;\;\;$ {\it Distance} between two neighbor points in the spatial frame:}  

\bea                                                                                  \label{dL}%02
\dd L:= \sqrt{(g_{0i}g_{0j}/g_{00}-g_{ij})\dd x^i\dd x^j}\,, \hspace{3mm} i,j=1,2,3\,.
\eea

\ppln{^Gi nuli\^gas se kaj nur se la punktoj koincidas.}%Atentu ke $\dd L$ 
\pprn{It is null if and only if the points coincide.}%See that $\dd L$

\ppln{$\;\;\;$ {\it Intertempo} do evento $x^\mu$ para o evento $x^\mu+\dd x^\mu$\,:}
\pprn{$\;\;\;$ {\it Intertime} (interval of time) from event $x^\mu$ to the event $x^\mu+\dd x^\mu$\,:} 

\bea                                                                                  \label{dT}%03
\dd T:= g_{0\mu}\,\dd x^\mu/(c\sqrt{g_{00}})\,.
\eea

\ppln{Se la intervalo $\dd x^\mu$ estas de tempa tipo (a\u u nula) kaj $\dd T$ estas pozitiva, tiuokaze objekto (a\u u lumo) povas iri de $x^\mu$ al $x^\mu+\dd x^\mu$\,. Se tamen $\dd T$ estas malpozitiva, la movado de objekto (a\u u lumo) estos de $x^\mu+\dd x^\mu$ al $x^\mu$\,. Se $\dd x^\mu$ estas spaca, $\dd T$ indikas la tempan ordon de eventoj por inercia observanto fiksata en la spaca teksa\jj o en $x^\mu$\,.}
\pprn{If the interval $\dd x^\mu$ is timelike (or null) and $\dd T$ is positive, then an object (or light) can go from event $x^\mu$ to the event $x^\mu+\dd x^\mu$\,. If however $\dd T$ is negative, the motion of the object (or light) will be from event $x^\mu+\dd x^\mu$ to the event $x^\mu$\,. If $\dd x^\mu$ is spacelike, then $\dd T$ gives the temporal order of the events for an inertial observer fixed in the spatial frame in $x^\mu$\,.}

\ppln{$\;\;\;$ Ni faru rimarkon pri propra intertempo. Atentu ke $(\dd s)^2$ estas difinata per linielemento~(\ref{SR1}), sed la signumo de $\dd s$ ne estas fiksata. Por $\dd s\neq0$\,, ni interkonsentas}
\pprn{$\;\;\;$ Let us make a remark about propertime interval. See that $(\dd s)^2$ is defined in the line element~(\ref{SR1}), but the signal of $\dd s$ is not fixed. For $\dd s\neq0$\,, we agree}

\bea                                                                                \label{dTds}%04
\dd T/\dd s\geq0\,.
\eea

\ppln{Speciale, en movadoj de tempa tipo $(\dd s)/c$ estas la intervalo $\dd\tau$ de pasita propratempo, do $s$ pligrandi\^gas dum la movado.}
\pprn{In particular, in the timelike motions $(\dd s)/c$ is the interval $\dd\tau$ of elapsed propertime, so $s$ increases in the motion.}

\ppln{$\;\;\;$ {\em Rapido} inter $x^\mu$ kaj $x^\mu+\dd x^\mu$\,:}
\pprn{$\;\;\;$ {\em Velocity} between $x^\mu$ and $x^\mu+\dd x^\mu$\,:} 

\bea                                                                                   \label{v}%05
V:=\dd L/|\dd T|\,.
\eea

\ppln{Oni konstatas ke $V<c$ se kaj nur se $\dd x^\mu$ estas de tempa tipo.}
\pprn{One finds that $V<c$ if and only if $\dd x^\mu$ is timelike.}

\ppln{$\;\;\;$ Difinoj (\ref{dL}) kaj (\ref{dT}) implicas}
\pprn{$\;\;\;$ The definitions (\ref{dL}) and (\ref{dT}) imply}

\bea                                                                                 \label{ds3}%06
\epsilon(\dd s)^2=(c\dd T)^2-(\dd L)^2\,,
\eea

\ppln{same kiel en speciala relativeco. Do $\dd T$\,, $\dd L$ kaj $V$ estas, respektive, intertempo, distanco kaj rapido, rilataj al movado de objekto (se $V<c$), en inercia referencsistemo restanta relative al la spaca teksa\jj o, en evento $x^\mu$\,; alivorte, inercia referencsistemo fiksata al spaca teksa\jj o en tiu punkto kaj en tiu momento.}
\pprn{exactly as in special relativity. So $\dd T$\,, $\dd L$ and $V$ are respectively time  interval, distance and velocity related to the motion of an object (if $V<c$), in an inertial reference system at rest relative to the spatial frame, in the event $x^\mu$\,; in other words, an inertial reference system fixed in the spatial frame in that point and in that moment.}

\ppln{$\;\;\;$ Por la metriko de Som-Raychaudhuri, la distanco kaj la intertempo estas}
\pprn{$\;\;\;$ For the metric of Som-Raychaudhuri, the distance and the intertime are}

\bea                                                                               \label{refut}%07
\dd L=\sqrt{(\dd r)^2+(r\dd\varphi)^2+(\dd z)^2}\,, \hspace{3mm} \dd T=\dd t-(\Omega r^2/c^2)\dd\varphi\,. 
\eea

\vspace{5mm}
\ppln{{\bf {\Large 2 \hspace{5mm}Ekvacioj de geodezioj}}}
\pprn{{\bf {\Large 2 \hspace{5mm}Equations of geodesics}}}
\ppln{} \pprn{}
%\ppsection[0.6ex]{Ekvacioj de geodezioj\label{Ekvacioj}}{\vspace{2cm}Equa\c c\~oes das geod\'esicas}                                                                             Sekcio 2
\ppln{Linielemento (\ref{SR1}) ne pendas de $t, \varphi, z$\,. Do la respondaj kovariantaj komponoj de la kvar\-rapido $u^\mu:=\dd x^\mu/\dd s$ estas konstantaj, en geodezia movado: $u_0=\eta,$ $u_\varphi=-\beta, \,u_z=-\zeta\,$. Uzante $u_\mu=g_{\mu\nu}u^\nu$, rezulti^gas}
\pprn{The line element (\ref{SR1}) does not depend on $t, \varphi, z$\,. So the corresponding covariant components of the fourvelocity $u^\mu:=\dd x^\mu/\dd s$ are constant, in a geodetic motion: $u_0=\eta,$ $u_\varphi=-\beta, \,u_z=-\zeta\,$. Using $u_\mu=g_{\mu\nu}u^\nu$, it results}

\bea                                                                                 \label{eta}%08
\eta&=&c\dd t/\dd s-(\Omega r^2/c)\dd\varphi/\dd s\,,  
\\                                                                                  \label{beta}%09
-\beta&=&(-\Omega r^2/c)(c\dd t/\dd s)-(r^2-\Omega^2r^4/c^2)\dd\varphi/\dd s\,,
\\                                                                               \label{eq.zeta}%10
-\zeta&=&-\dd z/\dd s\,.
\eea 

\ppln{Atentu ke (\ref{refut}b) kaj (\ref{eta}) implicas}
\pprn{See that (\ref{refut}b) and (\ref{eta}) imply}

\bea                                                                                \label{etaa}%11
\eta=c\dd T/\dd s\,.
\eea 

\ppln{Konsekvence, konvencio~(\ref{dTds}) implicas $\eta\geq0$\,.}
\pprn{Consequently, convention~(\ref{dTds}) implies $\eta\geq0$\,.}

\ppln{$\;\;\;$ Per rearan\^goj en (\ref{eta})--(\ref{eq.zeta}), okazas}
\pprn{$\;\;\;$ Rearranging (\ref{eta})--(\ref{eq.zeta}), results}

\bea                                                                                 \label{cdt}%12
c\dd t/\dd s&=&\beta\Omega/c+\eta(1-\Omega^2r^2/{c^2})\,, 
\\                                                                                   \label{dfi}%13
\dd\varphi/\dd s&=&\beta/r^2-\eta\Omega/c\,, 
\\                                                                                    \label{dz}%14
\dd z/\dd s&=&\zeta\,. 
\eea 

\ppln{Por havi $\dd r/\dd s$\,, ni metas (\ref{cdt})--(\ref{dz}) en lini\-elemento (\ref{SR1}), ricevante}
\pprn{To have $\dd r/\dd s$\,, we insert (\ref{cdt})--(\ref{dz}) in the line element (\ref{SR1}), obtaining}

\bea                                                                               \label{eq.dr}%15
(\dd r/\dd s)^2= \mu^2-r^2\left(\frac{\beta}{r^2}-\frac{\eta\Omega}{c}\right)^2\,, \hspace{3mm} \mu:=\sqrt{\eta^2-\zeta^2-\epsilon}\,. 
\eea

\ppln{\^Car $\mu^2=(\dd r/\dd s)^2+r^2(\dd\varphi/\dd s)^2$, $\mu$ estas nula nur en movado paralela al la akso $z$.}
\pprn{Since $\mu^2=(\dd r/\dd s)^2+r^2(\dd\varphi/\dd s)^2$, $\mu$ is null only in motions parallel to the $z$-axis .}

\ppln{$\;\;\;$ Nun ni komencas integri (\ref{cdt})--(\ref{eq.dr}). Unue ni kombinas (\ref{dfi}) kun (\ref{eq.dr}), ricevante}
\pprn{$\;\;\;$ Now we start integrating (\ref{cdt})--(\ref{eq.dr}). We first combine  (\ref{dfi}) with (\ref{eq.dr}), obtaining}

\bea                                                                                \label{drdf}%16
\left(\frac{\dd r}{\dd\varphi}\right)^2= \mu^2\left(\frac{\beta}{r^2}-\frac{\eta\Omega}{c}\right)^{-2}-r^2\,.
\eea

\ppln{La solvoj de (\ref{drdf}) estas cirkloj,}
\pprn{The solutions of (\ref{drdf}) are circles,}

\bea                                                                                \label{rho2}%17
r^2+a^2-2ar\cos(\varphi-\varphi_0)=\rho^2\,, \hspace{3mm} \rho:=\frac{c\mu}{2\eta\Omega}\,, \hspace{3mm} a:=\sqrt{\rho^2+\frac{c\beta}{\eta\Omega}}\,, 
\eea 

\ppln{kie $a$ kaj $\varphi_{\,0}$  estas respektive radiusa kaj angula pozicio de la cirkla centro, kaj $\rho$ estas la radiuso, kiel figuro~\ref{Baza} montras. \^Car $\mu$ en (\ref{eq.dr}b) ne pendas de $\beta$, anka\u u la radiuso $\rho$ ne pendas.}
\pprn{where $a$ and $\varphi_{\,0}$  are respectively the radial and angular positions of the center of the circle, and $\rho$ is the radius, as figure~\ref{Baza} shows. Since $\mu$ in (\ref{eq.dr}b) does not depend on $\beta$, neither the radius $\rho$ depends.}

\begin{figure}[h]                                                                        %Figura 2
\centerline{\epsfig{file=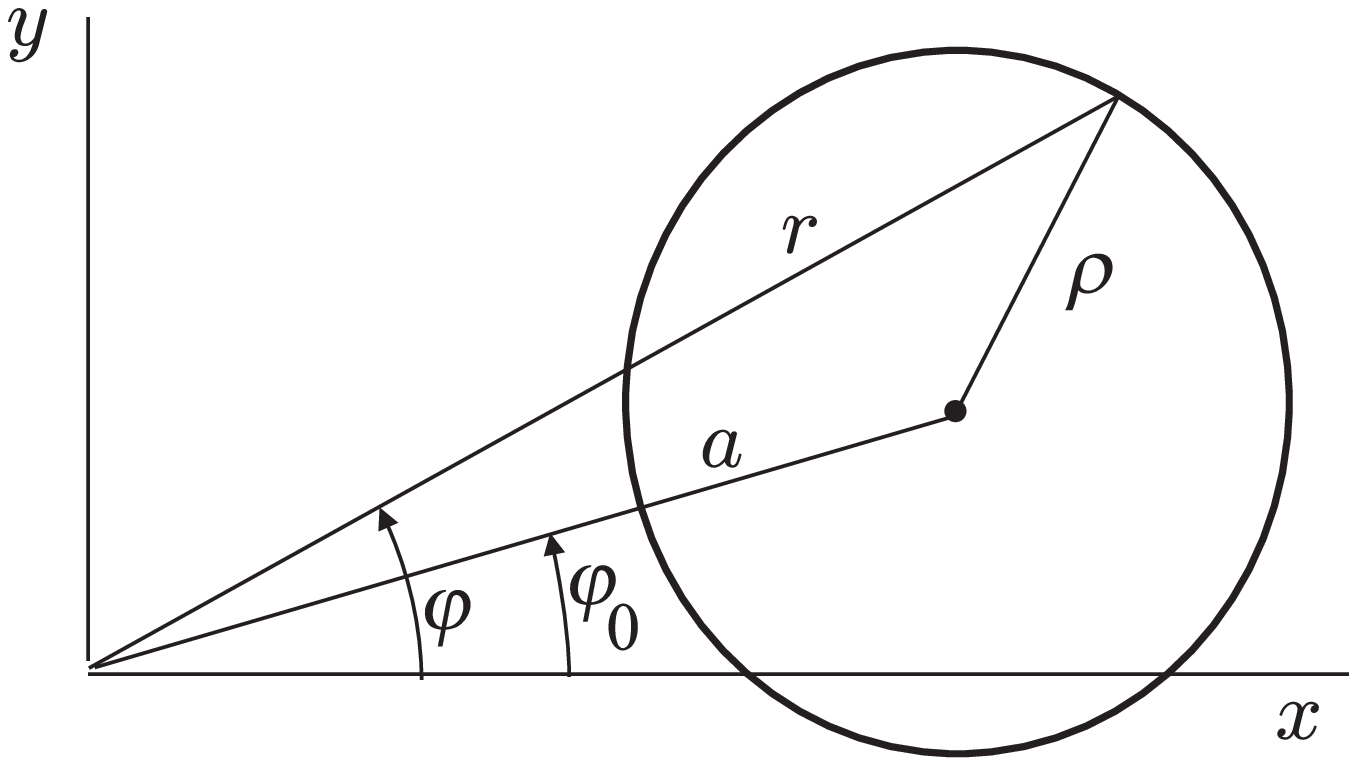,width=7cm}} 
\caption{
La cirklo (\ref{rho2}) por okazo $a>\rho$\,. La cirklo estas orta projekcio de geodezia trajektorio, en ebeno $z={\rm konst}$.
\newline 
Figura~\ref{Baza}: 
The circle (\ref{rho2}) for the case $a>\rho$\,. It is the orthogonal projection of the geodetic trajectory in a plane $z={\rm const}$.}
\label{Baza} 
\end{figure}

\ppln{$\;\;\;$ Do \^ciu geodezia trajektorio estas desegnata en cirkla cilindro kun radiuso $\rho$ kaj akso paralela al akso $z$, je distanco $a$\,. Tio estas montrata en figuro~\ref{Baza} por $a>\rho$, ekvivalente $\beta>0$\,. Figuro~\ref{Cirkloj} montras aliajn okazojn, por kelkaj valoroj de  $\beta$\,. Se $\beta=0$\,, tial $a=\rho$, kaj la trajektorio krucas la akson $z$\,. Se $\beta<0$\,, la cilindro ^cirka\u ufermas la akson. En speciala okazo $\beta=-c\mu^2/(4\eta\Omega)$, okazas $a=0$\,, tio estas, la akso de cilindro estas la akso $z$\,.}
\pprn{$\;\;$ So every geodetic trajectory is drawn in a circular cylinder with radius $\rho$ and axis parallel to the $z$-axis , at a distance $a$\,. This is shown in figure~\ref{Baza} for $a>\rho$, equivalently $\beta>0$\,. Figure~\ref{Cirkloj} shows other cases, for some values of $\beta$\,. If $\beta=0$\,, then $a=\rho$, and the trajectory crosses the $z$-axis . If $\beta<0$\,, the cylinder encircles the axis. In the special case $\beta=-c\mu^2/(4\eta\Omega)$, it occurs $a=0$\,, that is, the axis of the cylinder is the $z$-axis .}

\ppln{$\;\;\;$ Oni konstatas ke la kalkuloj de geodezioj kun $a=0$ estas tre simplaj. Ili estas faritaj en la sekvanta sekcio. \^Car \cite{PhysLett} montris ke la linielemento (\ref{SR1}) estas spactempe homogena, ni povas forkonduki geodeziojn kun $a=0$ al iu regiono de spactempo. Sekcio~4 studos malkune la interesan okazon $a\rightarrow\infty$.}
\pprn{$\;\;$ One finds that calculation of geodesics with $a=0$ are very simple. They are made in the next section. Since \cite{PhysLett} shew that the line element (\ref{SR1}) is homogeneous in all spacetime, we can carry geodesics with $a=0$ to any region of spacetime. Section~4 will study separately the interesting case $a\rightarrow\infty$.}

\begin{figure}[ht]                                                                        %Figura 3
\centerline{\epsfig{file=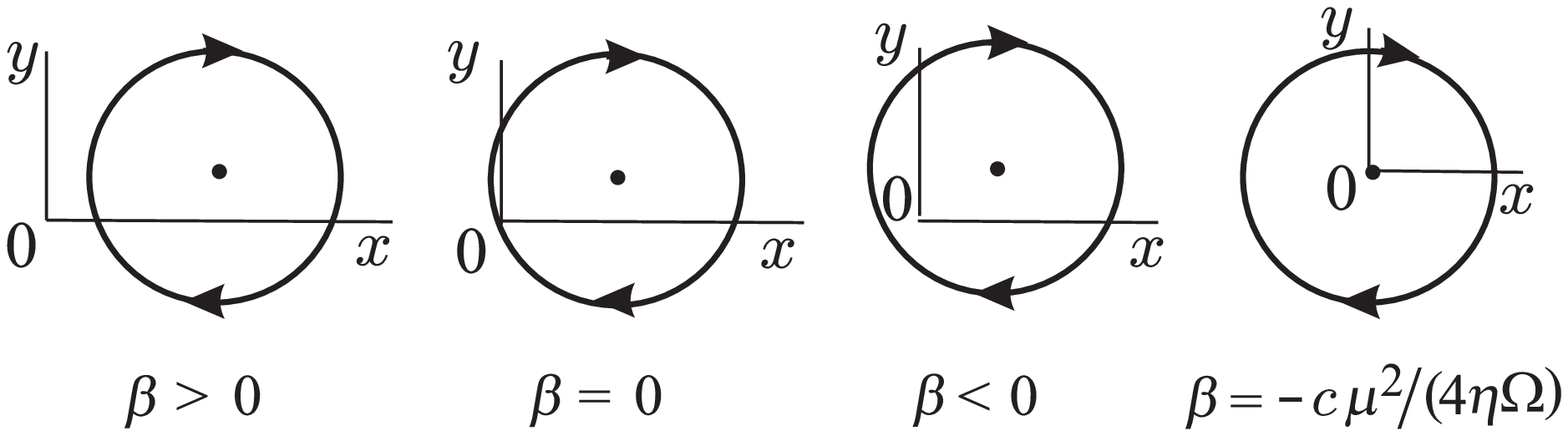,width=11cm}}
\caption{Orta sekcio de cirkla cilindro subtenante geodezian trajektorion. La pozicio de la cilindra akso relative al akso $z$ (tra origino 0) pendas de $\beta$\,. Sago indikas direkton de pligrandi\^go de $T$ en geodezio. 
\newline 
Figura~\ref{Cirkloj}: Orthogonal section of the circular cylinder that bears a geodetic trajectory. The position of cylinder's axis relative to the $z$-axis  (through origin 0) depends on $\beta$\,. Arrows show the direction of increase of $T$ in the geodesic.}
\label{Cirkloj} 
\end{figure}

\vspace{5mm}
\ppln{{\bf {\Large 3 \hspace{5mm}Geodezio kun $a=0$}}}
\pprn{{\bf {\Large 3 \hspace{5mm}Geodesic with $a=0$}}}
\ppln{} \pprn{}
%\ppsection[0.6ex]{Geodezioj kun $a=0$ \label{anula}}{Geod\'esicas com $a=0$}              Sekcio 3
\ppln{Farante $a=0$\,, $r=\rho={\rm konst}$, kaj $\beta=-c\mu^2/(4\eta\Omega$)\, en (\ref{cdt})--(\ref{dz}), ni ricevas}
\pprn{Setting $a=0$\,, $r=\rho={\rm const}$, and $\beta=-c\mu^2/(4\eta\Omega$)\, in (\ref{cdt})--(\ref{dz}), we get} 

\bea                                                                               \label{novas}%18
\frac{c\dd t}{\dd s}=\frac{\eta^2+\zeta^2+\epsilon}{2\eta}\,, \hspace{6mm} \frac{\dd\varphi}{\dd s}=-\frac{2\eta\Omega}{c}\,, \hspace{6mm} \frac{\dd z}{\dd s}=\zeta\,. 
\eea

\ppln{\^Car la kvociento $\dd z/\dd\varphi$ estas konstanta, ni konstatas ke la pa\^so de la cirkla helica trajektorio estas konstanta.}
\pprn{Since the quocient $\dd z/\dd\varphi$ is constant, we see that the pitch of the circular helical trajectory is a constant.}

\ppln{$\;\;\;$ La malpozitiva signumo en (\ref{novas}b) implicas ke la geodezia parametro $s$ pligrandi\~gas en la direkto anti-$\Omega$. Do (\ref{dTds}) diras ke la tempa parametro $T$ anka\u u pligrandi\^gas en la direkto anti-$\Omega$. Tio estas jam indikita en figuro~\ref{Cirkloj}.}
\pprn{$\;\;\;$ The negative sign in (\ref{novas}b) implies that the geodetic parameter $s$ increases in the anti-$\Omega$ direction. Then (\ref{dTds}) says that also the time parameter $T$ increases in the anti-$\Omega$ direction. This was already indicated in figure~\ref{Cirkloj}.}

\ppln{$\;\;\;$ Ni integras (\ref{novas}b) farante $\varphi$ varii de 0 al $-2\pi$ kaj ricevas $\Delta s$\,, la pligrandi\^gon de $s$ dum unu kompleta helicero. Poste, uzante~(\ref{novas}a) kaj (\ref{novas}c) ni ricevas $\Delta t$ kaj $\Delta z$\, akumulitaj en unu helicero:}
\pprn{$\;\;\;$ We integrate (\ref{novas}b) with $\varphi$ varying from 0 to $-2\pi$ and get $\Delta s$\,, the increase of $s$ in a complete spire. Then, using~(\ref{novas}a) and (\ref{novas}c) we find $\Delta t$ and $\Delta z$\, accumulated in one complete spire:} 

\bea                                                                              \label{novas2}%19
\Delta\varphi=-2\pi\,,
\hspace{4mm} \Delta s=\frac{\pi c}{\eta\Omega}\,, 
\hspace{4mm} \Delta t= \frac{\pi}{2\Omega}\left(1+\frac{\epsilon+\zeta^2}{\eta^2}\right)\,, \hspace{4mm} \Delta z=\frac{\pi\zeta c}{\eta\Omega}\,.
\eea

\ppln{Por kompleti la priskribon de \^ci tiuj geodezioj, ni uzas~(\ref{rho2}b) por ricevi la radiuson de la cirkla helico,}
\pprn{To complete the description of these geodesics, we use~(\ref{rho2}b) to have the radius of the circular helix,}

\bea                                                                            \label{novasrho}%20
\rho=\frac{c}{2\Omega}\sqrt{1-\frac{\epsilon+\zeta^2}{\eta^2}}\,.
\eea

\ppln{$\;\;\;$ Por prezenti pli klaran priskribon, ni substituas parametrojn $\eta$ kaj $\zeta$ kun aliaj pli familiaraj. Uzante~(\ref{ds3}) kaj~(\ref{etaa}) ni ricevas la rapidon $V$, difinita en~(\ref{v}), kiel funkcio de $\eta$\,, kaj \^gian inverson, montritan en figuro~\ref{Eta}:}
\pprn{$\;\;\;$ In order to present a more clear description, we replace the parameters $\eta$ and $\zeta$ by other more familiar. Using~(\ref{ds3}) and~(\ref{etaa}) we obtain the velocity $V$, defined in~(\ref{v}), as a function of $\eta$\,, and its inverse, shown in figure~\ref{Eta}:}

\bea                                                                                \label{defv}%21
V/c=\sqrt{1-\epsilon/\eta^2}\,, \hspace{3mm} 
\eta =\frac{1}{\sqrt{\epsilon(1-V^2/c^2)}}\,;
\eea

\ppln{atentu ke $V$ estas konstanta.}
\pprn{see that $V$ is constant.}

\begin{figure}[ht]                                                                        %Figura 4
\centerline{\epsfig{file=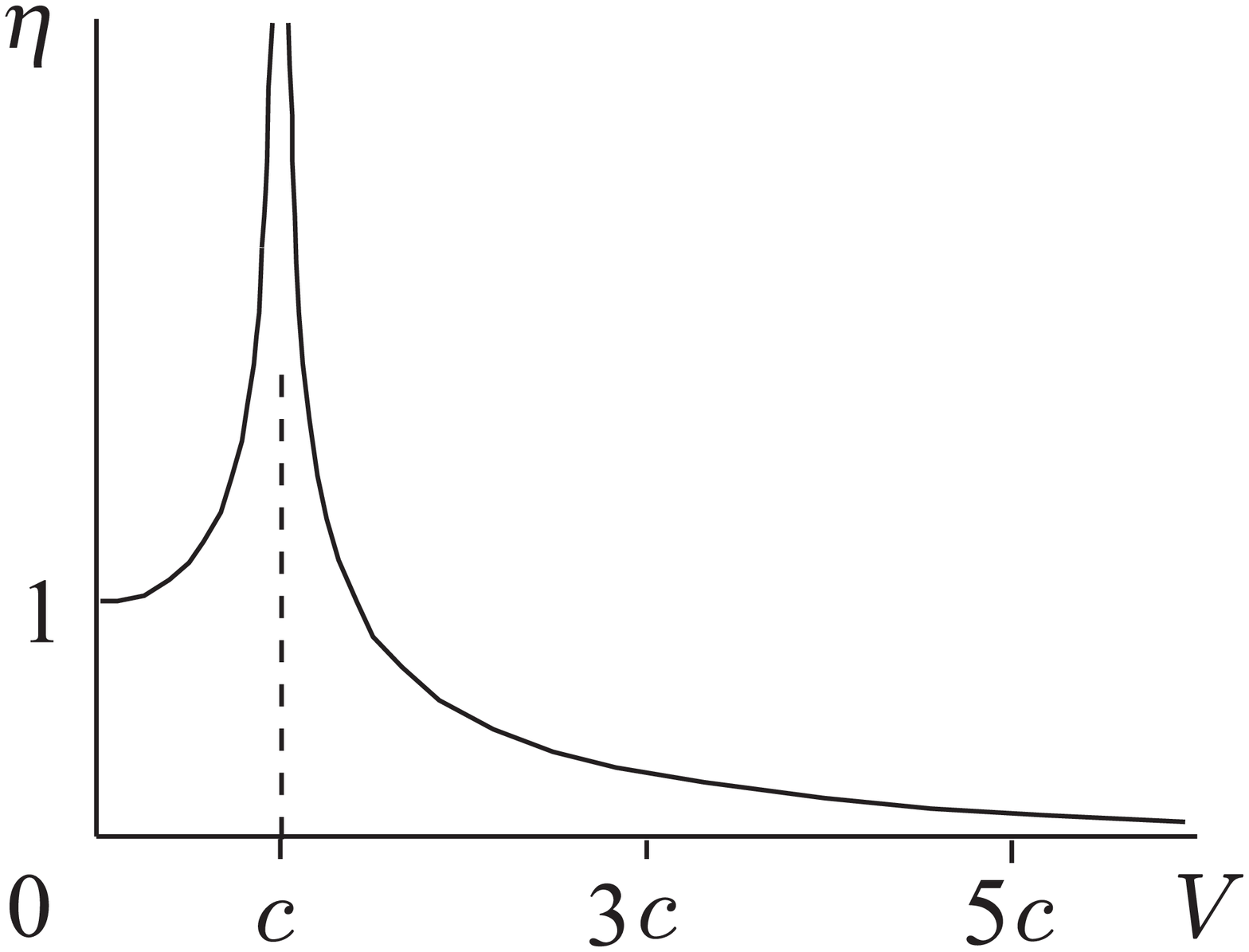,width=5cm}}
\caption{Rilato (\ref{defv}) inter parametroj $\eta$ kaj $V$\,.
\newline
Figura~\ref{Eta}: Relation (\ref{defv}) between parameters $\eta$ and $V$\,.}
\label{Eta} 
\end{figure}

\ppln{$\;\;\;$ Ni uzas anka\u u la angulon $\alpha\in[-\pi/2,$ $\pi/2]$\,, inter la orientita trajektorio kaj ebenoj $z={\rm konst}$:}
\pprn{$\;\;$ We use also the angle $\alpha\in[-\pi/2,$ $\pi/2]$\,, between the directed trajectory and planes $z={\rm const}$:}

\bea                                                                        \label{eq'tan'alpha}%22
\tan\alpha:=\dd z/\sqrt{(\dd r)^2+(r\,\dd\varphi)^2}\,. 
\eea

\ppln{\^Car \^ci tie $r={\rm konst}$, tial \mbox{$\tan\alpha:=\dd z/|\rho\dd\varphi|$}. Uzante la geodeziajn ekvaciojn (\ref{novas}b,c), kaj (\ref{novasrho})--(\ref{eq'tan'alpha}), ni ricevas}
\pprn{Since here $r={\rm const}$, then \mbox{$\tan\alpha:=\dd z/|\rho\dd\varphi|$}. Using the geodetic equations (\ref{novas}b,c), and (\ref{novasrho})--(\ref{eq'tan'alpha}), we get}

\bea                                                                               \label{alpha}%23
\zeta=\frac{\eta V}{c}\sin\alpha\,.
\eea 

\ppln{$\;\;\;$ Do, uzante parametroj $V$ kaj $\alpha$, la varioj~(\ref{novas2}) en unu kompleta helicero kaj la radiuso~(\ref{novasrho}) reskribi\^gas}
\pprn{$\;\;$ So, using parameters $V$ and $\alpha$, the variations~(\ref{novas2}) in one complete spire and the radius~(\ref{novasrho}) rewrite}

\bea                                                                              \label{novas3}%24
\Delta\varphi=-2\pi\,,
\hspace{4mm} \Delta s=\frac{\pi c}{\Omega}\sqrt{\epsilon(1-V^2/c^2)}\,,
\hspace{4mm}
\Delta t=\frac{\pi}{\Omega}(1-\frac{V^2}{2c^2}\cos^2\alpha)\,,
\hspace{4mm} \Delta z=\frac{\pi V}{\Omega}\sin\alpha\,,
\eea
\bea                                                                           \label{novas3rho}%25
\rho=\frac{V}{2\Omega}\cos\alpha\,. 
\eea

\ppln{$\;\;\;$ Ekvacioj~(\ref{novas3rho}) kaj~(\ref{novas3}d) diras ke la radiuso $\rho$ kaj la pa\^so $|\Delta z|$ de helico pendas lineare de la konstanta rapido $V$\,; sed $\rho$ plieti\^gas se $|\alpha|$ pligrandi\^gas, kvankam $|\Delta z|$ pligrandi\^gas. Integrante (\ref{refut}a) por unu kompleta helicero, poste uzante ekvaciojn~(\ref{novas}) de geodezioj kaj~(\ref{alpha}) kaj~(\ref{novas3rho}), ni ricevas}
\pprn{$\;\;\;$ Equations~(\ref{novas3rho}) and~(\ref{novas3}d) say that the radius $\rho$ and helix's pitch $|\Delta z|$ depend linearly on the constant velocity $V$\,; but $\rho$ shrinks if  $|\alpha|$ increases, while $|\Delta z|$ increases. Integrating (\ref{refut}a) for a complete spire, then using equations (\ref{novas}) of the geodesics and (\ref{alpha}) and~(\ref{novas3rho}), we obtain}

\bea                                                                              \label{DeltaL}%26
\Delta L=\frac{\pi V}{\Omega}\,, 
\eea 

\ppln{montrante ke la longo de helicero ne pendas de la klino $\alpha$\,. Do trajektorio similas  risorton. Fakte, premante risorton, la pa\^so $|\Delta z|$ plieti\^gas proporcie al pligrandi\^go de radiuso $\rho$\,. Tio permesas facile konstrui geodeziajn trajektoriojn kun sama proporcio $V/\Omega$\,, kiel figuro~\ref{Helice} montras.}
\pprn{showing that the lenght of a spire does not depend on the slope $\alpha$\,. So a trajectory ressembles a spring. Really, if we press a spring, the pitch $|\Delta z|$ shrinks proportionally to the increase of the radius $\rho$\,. This makes easy to construct geodetic trajectories with same ratio $V/\Omega$\,, as figure~\ref{Helice} shows.}

\begin{figure}[ht]                                                                        %Figura 5
\centerline{\epsfig{file=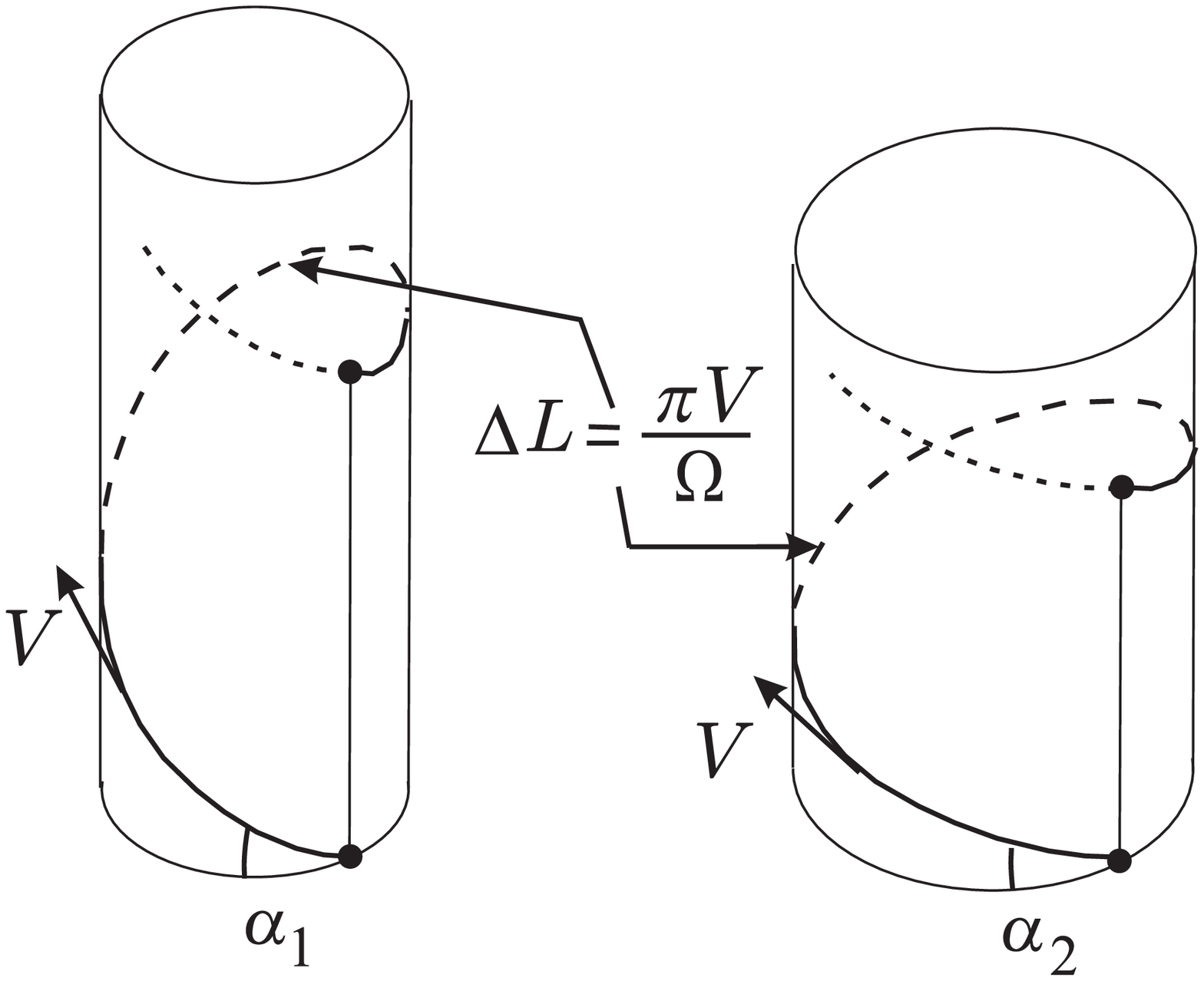,width=6cm}} 
\caption{
Heliceroj kun sama proporcio $V/\Omega$ havas la saman longon.
\newline 
Figura~\ref{Helice}: Spires with same ratio $V/\Omega$ have the same lenght.}
\label{Helice} 
\end{figure}

\ppln{$\;\;\;$ Ni kalkuku anka\u u la intertempon $|\Delta T|$ en unu helicero. \^Car $V$ estas konstanta, tial $|\Delta T|= \Delta L/V$\,, tio estas,}
\pprn{$\;\;$ We calculate also the intertime $|\Delta T|$ within one spire. Since $V$ is constant, then $|\Delta T|= \Delta L/V$\,, that is,} 

\bea                                                                              \label{DeltaT}%27
|\Delta T|=\frac{\pi}{\Omega}\,. 
\eea 

\ppln{Ni vidas ke $|\Delta T|$ pendas nek de $V$ nek de $\alpha$\,. Vere, $|\Delta T|$ estas la duono de la periodo $2\pi/\Omega$ de loka rotacio de spaca teksa\jj o relative al la inercia kompaso.}
\pprn{We see that $|\Delta T|$ depends neither on $V$ nor on $\alpha$\,. Really, $|\Delta T|$ is half the period $2\pi/\Omega$ of local rotation of the spatial frame relative to the compass of inertia.} 

\ppln{$\;\;\;$ En la tri sekvantaj subsekcioj, ni konsideras malkune geodeziojn de tempa tipo, nula tipo, kaj spaca tipo.}
\pprn{$\;\;\;$ In the next three subsections we consider separately the timelike, null, and spacelike geodesics.}

\newpage%\vspace{5mm}
\ppln{{\bf {\large 3.1 \hspace{5mm}Geodezio de tempa tipo}}}
\pprn{{\bf {\large 3.1 \hspace{5mm}Timelike geodesic}}}
\ppln{} \pprn{}
%\ppsubsection[0.6ex]{Geodezioj de tempa tipo \label{tempaj}}{Geod\'esicas tipo tempo}Subsekcio 3.1
\ppln{Ekvacio~(\ref{novas3rho}) montras ke la radiuso $\rho$ de helico estas proporcia al rapido $V$, kaj $\rho$ plieti\^gas se la klino $\alpha$ pligrandi\^gas. Plu, $\rho<c/(2\Omega)$. En \^ci tiu limo, la movado estas plana. Aliflanke, se la klino estas maksimuma ($|\alpha|=\pi/2$), do la radiuso $\rho$ estas nula, kaj la movado okazas en akso $z$\,. Tiuokaze, se $\zeta=0$ do $V=0$, indikante objekton restantan en akso $z$\,. Kiel ni klarigis en sekcio~2, ni povas forkonduki \^ci tiun geodezion al iu regiono de spacotempo, do konstruante la geodeziojn de la polvo generanta graviton.}
\pprn{Equation~(\ref{novas3rho}) shows that the radius $\rho$ of the helix is linearly proportional to the velocity $V$, and is shorter if the slope $\alpha$ is greater. Further, $\rho<c/(2\Omega)$. In this limit, the motion is planar. On the other side, if the slope is maximum ($|\alpha|=\pi/2$), then the radius $\rho$ is null and the motion is along the $z$-axis. In this case, if $\zeta=0$ then $V=0$, indicating an object at rest on the $z$-axis. As we clarified in section~2, we can transport this geodesic to any region of spacetime, thus constructing the geodesics of the dust generating gravitation.}

\ppln{$\;\;\;$ Ekvacio~(\ref{novas3}c) montras ke la intertempo $\Delta t$ en unu kompleta helicero estas pozitiva, kaj valoras inter $\pi/(2\Omega)$ kaj $\pi/\Omega$\,; \^gi estas ju pli granda des malpli la rapido $V$ estas granda, kaj des pli la trajektorio estas vertikala. Ni konstatas ke (\ref{novas3}c) malpermesas ka\u uzan malobeon en geodezioj de tempa tipo kaj nula tipo; ili \^ciam havas $\Delta t>0$\,.}
\pprn{$\;\;\;$ Equation~(\ref{novas3}c) shows that the interval $\Delta t$ of time coordinate in one complete spire is always positive, with value between $\pi/(2\Omega)$ and $\pi/\Omega$\,; it is larger if the velocity $V$ is shorter, and if the trajectory is more vertical. We see that (\ref{novas3}c) does not allow violation of causality in timelike and in null geodesics: these have always $\Delta t>0$\,.}

\ppln{$\;\;\;$ En geodezioj de tempa tipo kun $r={\rm konst}$, (\ref{cdt}) indikas ke $\dd t/\dd s$ estas konstanta. Sed en geodezioj kun $r\neq{\rm konst}$\,, $\dd t/\dd s$ ne nur estas malkonstanta, sed estas malpozitiva en regionoj kun $r$ sufi\^ce granda. Tio signifas ke voja\^ganto kun tiu movado renkontas valorojn de $t$ sinsekve plietaj en tiuj regionoj. Tamen, en unu kompleta helicero, la pozitivaj kontribuoj de $\dd t/\dd s$ en regionoj kun malgrandaj $r$ superas tiujn malpozitivajn kontribuojn, okazante pozitivan rezulton por $\Delta t$.}
\pprn{$\;\;\;$ In timelike geodesics with $r={\rm const}$, equation (\ref{cdt}) indicates that $\dd t/\dd s$ is constant. But in geodesics with $r\neq{\rm const}$, $\dd t/\dd s$ not only varies, but is negative in regions with $r$ suficiently large. This means that a voyager with that motion finds values of local $t$ successively shorter in these regions. However, in a complete spire, the positive contributions of $\dd t/\dd s$ in regions with shorter $r$ surpass the negative contributions from larger $r$, producing a positive result for $\Delta t$.}

\ppln{$\;\;\;$ La pa\^so $|\Delta z|$ de helico en (\ref{novas3}d) estas proporcia al la rapido $V$, kaj estas pli granda se la trajektorio estas pli vertikala; \^gia maksimuma valoro estas $\pi c/\Omega$.}
\pprn{$\;\;\;$ The pitch $|\Delta z|$ of the helix in (\ref{novas3}d) is linearly proportional to the velocity $V$, and is greater if the trajectory is more vertical; its maximum value is $\pi c/\Omega$.}

\ppln{$\;\;\;$ La propra intertempo $\Delta\tau:=(\Delta s)/c$ en unu kompleta helicero estas en (\ref{novas3}b):}
\pprn{$\;\;\;$ The proper intertime $\Delta\tau:=(\Delta s)/c$ in a complete spire is in (\ref{novas3}b):}

\bea                                                                             \label{eq.Dtau}%28
\Delta\tau=\frac{\pi}{\Omega}\sqrt{1-V^2/c^2}\,.
\eea

\ppln{Ni konstatas ke $\Delta\tau$ ne pendas de la klino~$\alpha$\,. Anka\u u ke, en unu kompleta helicero, la propra intertempo~(\ref{eq.Dtau}) estas plieta ol la intertempo~(\ref{DeltaT}). Efektive, $(c\Delta\tau)^2=(c\Delta T)^2-(\Delta L)^2$\,.}
\pprn{We see that $\Delta\tau$ does not depend on the slope~$\alpha$\,. We see also that, in a complete spire, the proper intertime~(\ref{eq.Dtau}) is shorter than  the intertime~(\ref{DeltaT}). Indeed, $(c\Delta\tau)^2=(c\Delta T)^2-(\Delta L)^2$\,.}

\newpage%\vspace{5mm}
\ppln{{\bf {\large 3.2 \hspace{5mm}Nula geodezio}}}
\pprn{{\bf {\large 3.2 \hspace{5mm}Null geodesic}}}
\ppln{} \pprn{}
%\ppsubsection{Nulaj geodezioj \label{nulaj}}{Geod\'esicas nulas}                    Subsekcio 3.2
\ppln{Se $V=c$\,, (\ref{novas3})--(\ref{DeltaT}) simpli\^gas al}
\pprn{If $V=c$\,, (\ref{novas3})--(\ref{DeltaT}) simplify to}

\bea                                                                               \label{nulas}%29
\Delta t=\frac{\pi}{2\Omega}(1+\sin^2\alpha)\,, 
\hspace{3mm} 
\Delta z=\frac{\pi c}{\Omega}\sin\alpha\,,
\hspace{3mm} 
\rho=\frac{c}{2\Omega}\cos\alpha\,, 
\hspace{3mm} 
\Delta L=\frac{\pi c}{\Omega}\,, 
\hspace{3mm} 
|\Delta T|=\frac{\pi}{\Omega}\,. 
\eea

\ppln{Ni konstatas ke la konstanta radiuso (\ref{nulas}c) de luma helico estas malgranda se la klino $|\alpha|$ estas granda. Speciale, se $|\alpha|=\pi/2$, tial $\rho=0$, kaj la movado de lumo estas en akso $z$. Aliflanke, la maksimuma diametro de helico estas $2\rho=c/\Omega$\,, okazanta se la movado estas plana. Do la horizonto de eventoj por iu observanto estas cilindra surfaco kun radiuso $c/\Omega$\,, kies akso estas paralela al akso $z$ kaj krucas la observanto~\cite{PhysLett}.}
\pprn{We see that the constant radius (\ref{nulas}c) of the light helix is short if the slope  $|\alpha|$ is large. In particular, if $|\alpha|=\pi/2$, then $\rho=0$, and the motion of light is along the $z$-axis . On the other side, the maximum diameter of a helix is $2\rho=c/\Omega$\,, occurring if the motion is planar. So the event horizon for any observer is a cylindrical surface with radius $c/\Omega$\,, whose axis is parallel to the $z$-axis  and crosses the observer~\cite{PhysLett}.}

\ppln{$\;\;\;$ Ekvacioj (\ref{nulas}d) kaj(\ref{nulas}e) prezentas la longon de luma helicero, kaj la intertempon por trakuri ^gin. Fine, (\ref{nulas}a) montras ke en nulaj geodezioj anka\u u ne estas kauza malobeo.}
\pprn{$\;\;\;$ Equations (\ref{nulas}d) and (\ref{nulas}e) give the lenght of a lightspire, and the intertime to traverse it. Finally, (\ref{nulas}a) shows that neither in the null geodesics there is violation of causality.}

\vspace{5mm}
\ppln{{\bf {\large 3.3 \hspace{5mm}Geodezio de spaca tipo}}}
\pprn{{\bf {\large 3.3 \hspace{5mm}Spacelike geodesic}}}
\ppln{} \pprn{}
%\ppsubsection[0.6ex]{Geodezioj de spaca tipo \label{spacaj}}{Geod\'esicas tipo espa\c co} Subs.3.3
\ppln{Simile kiel en tempa kaj nula tipoj, la helicaj movadoj de spaca tipo povas i^gi rektaj, en akso $z$. Aliflanke, ne estas limesa supremo por radiuso $\rho$, kiu estas proporcia al rapido $V$. Ekvacioj (\ref{novas3}) kaj (\ref{novas3rho}) permesas spactempe fermitajn geodeziojn ($\Delta t=0$). Ili anka\u u permesas $\Delta t<0$, sed tio ne implicas kauzan malobeon, \^car kurbo de spaca tipo ne respondas al movado de materio nek de lumo.}
\pprn{Similarly as in the timelike and null motions, the spacelike helical can get straight,on the $z$-axis . On the other side, there is no upper limit for the radius $\rho$, which is linearly proportional to the velocity $V$. Equations~(\ref{novas3}) and~(\ref{novas3rho}) allow closed geodesics in spacetime ($\Delta t=0$). They also allow $\Delta t<0$, but this does not constitute causality violation, since a spacelike curve does not correspond to motion of matter or light.}

\ppln{$\;\;\;$ La propra longo $\Delta\lambda:=\Delta s$ de helicero de spaca tipo estas en (\ref{novas3}b), kun $\epsilon=-1$:}
\pprn{$\;\;\;$ The properlenght $\Delta\lambda:=\Delta s$ of a spacelike spire is in (\ref{novas3}b), with $\epsilon=-1$:} 

\bea                                                                             \label{Dlambda}%30
\Delta\lambda=\frac{\pi c}{\Omega}\sqrt{V^2/c^2-1}\,. 
\eea 

\ppln{Tio estas plieta ol la longo $\Delta L=\pi  V/\Omega$ de helicero. Efektive, $(\Delta\lambda)^2=(\Delta L)^2-(c\Delta T)^2$\,.}
\pprn{This is shorter than the lenght $\Delta L=\pi  V/\Omega$ of the spire. Indeed, $(\Delta\lambda)^2=(\Delta L)^2-(c\Delta T)^2$\,.}

\newpage%\vspace{5mm}
\ppln{{\bf {\Large 4 \hspace{5mm}Samtempa geodezio}}}
\pprn{{\bf {\Large 4 \hspace{5mm}Geod. of simultaneities}}}
\ppln{} \pprn{}
%\ppsection[0.6ex]{Samtempaj geodezioj \label{samtempaj}}{Geod\'esicas de simulta-\\ neidades}Sek.4
\ppln{En elektata koordinatsistemo, samtempa geodezio estas tiu de spaca tipo $(\epsilon=-1)$, kies intervaloj $\dd x^\mu$ havas intertempon $\dd T$ nulan \cite{PaivaTeixeira2010s,reltemp2}, do rapido $V\rightarrow\infty$\,. Atentante~(\ref{eta}) oni vidas ke $\eta\rightarrow0$\,, kaj $V\eta/c\rightarrow1$, kaj do (\ref{rho2}) implicas radiuson $\rho\rightarrow\infty$ kaj distancon $a\rightarrow\infty$\,. Tamen, la troa\jj o $|a-\rho|$ povas esti limhava, kiel ni supozas en \^ci tiu sekcio.}
\pprn{In a given coordinate system, geodesic of simultaneities is a spacelike geodesic $(\epsilon=-1)$ whose infinitesimal intervals $\dd x^\mu$ have intertime $\dd T$ null  \cite{PaivaTeixeira2010s,reltemp2}, and so velocity $V\rightarrow\infty$\,. Remembering~(\ref{eta}) one sees that $\eta\rightarrow0$\,, and $V\eta/c\rightarrow1$, so equations~(\ref{rho2}) imply radius $\rho\rightarrow\infty$ and distance $a\rightarrow\infty$\,. However, the difference $|a-\rho|$ can be finite, as we assume in this section.}

\ppln{$\;\;\;$ \^Car $a\neq0$, la rezultoj de sekcio~3 ne validas \^ci tie. Do ni reiras al sekcio~2. Uzante difinon de $\alpha$ en (\ref{eq'tan'alpha}), kaj uzante (\ref{dfi})--(\ref{eq.dr}), ni ricevas $\zeta=\sin\alpha$ kaj $\mu=\cos\alpha$\,. Difinante}
\pprn{$\;\;\;$ Since $a\neq0$, the results of section~3 do not apply here. We then come back to section~2. Using $\alpha$ defined in (\ref{eq'tan'alpha}), and using (\ref{dfi})--(\ref{eq.dr}), we obtain $\zeta=\sin\alpha$ and  $\mu=\cos\alpha$\,. Defining}

\bea                                                                                   \label{D}%31
D:=\beta/\cos\alpha\,, 
\eea

\ppln{ni reskribas (\ref{cdt})--(\ref{eq.dr}) kiel}
\pprn{we rewrite (\ref{cdt})--(\ref{eq.dr}) as}

\bea                                                                               \label{simul}%32
\frac{\dd t}{\dd s}=\frac{D\Omega}{c^2}\cos\alpha\,, \hspace{5mm} \frac{\dd\varphi}{\dd s}=\frac{D}{r^2}\cos\alpha\,, \hspace{5mm} \frac{\dd z}{\dd s}=\sin\alpha\,, \hspace{5mm} \left(\frac{\dd r}{\dd s}\right)^2=(1-D^2/r^2)\cos^2\alpha\,.
\eea

\ppln{\^Car $\dd T=0$ en (\ref{ds3}) implicas $|\dd s|=\dd L$\,, ekvacioj (\ref{simul}b,c,d) estas la bonkonataj ekvacioj de rekto en trispaco.}
\pprn{Since $\dd T=0$ in (\ref{ds3}) implies $|\dd s|=\dd L$\,, the equations (\ref{simul}b,c,d) are the well known equations of a straight line in threespace.} 

\ppln{$\;\;\;$ Fakte, ni kombinas~(\ref{simul}b) kun~(\ref{simul}d):}
\pprn{$\;\;\;$ Indeed, we combine~(\ref{simul}b) with~(\ref{simul}d):}

\bea                                                                                \label{dfdr}%33
\left(\dd\varphi/\dd r\right)^2=\frac{1}{r^2(r^2/D^2-1)}\,,
\eea

\ppln{kies solvo estas}
\pprn{whose solution is}

\bea                                                                                \label{reta}%34
r\cos(\varphi-\varphi_0)=D\,, \hspace{3mm} \varphi_0={\rm const}\,.
\eea

\ppln{Kiel figuro~\ref{Reta} montras, tiu solvo estas rekto kies distanco al akso $z$ estas $|D|$. Tiu rekto estas projekcio de geodezia trajektorio en ebenoj $z=$ konst.}
\pprn{As figure~\ref{Reta} shows, this solution is a straight line whose distance to the $z$-axis  is $|D|$\,. That line is the projection of the geodetic trajectory onto planes $z=$ const.} 

\ppln{$\;\;\;$ Ekvacio~(\ref{simul}a) informas ke la vario de tempa koordinato $t$ estas unuforma, la\u ulonge la trajektorio. Kaj informas ke la lega\jj oj de $t$ ne varias, en movadoj paralelaj al akso $z$ (tiuokaze $\alpha=\pm\pi/2$). Fine, komparante (\ref{simul}a) kun (\ref{simul}b), ni konstatas ke $\dd\varphi/\dd t>0$.}
\pprn{$\;\;\;$ Equation~(\ref{simul}a) says that the variation of the time coordinate $t$ is uniform, along the trajectory. It also says that the readings of $t$ do not change, in motions parallel to the $z$-axis (cases where $\alpha=\pm\pi/2$). Finally, comparing (\ref{simul}a) with (\ref{simul}b), we see that $\dd\varphi/\dd t>0$.}

\begin{figure}[ht]                                                                        %Figura 6
\centerline{\epsfig{file=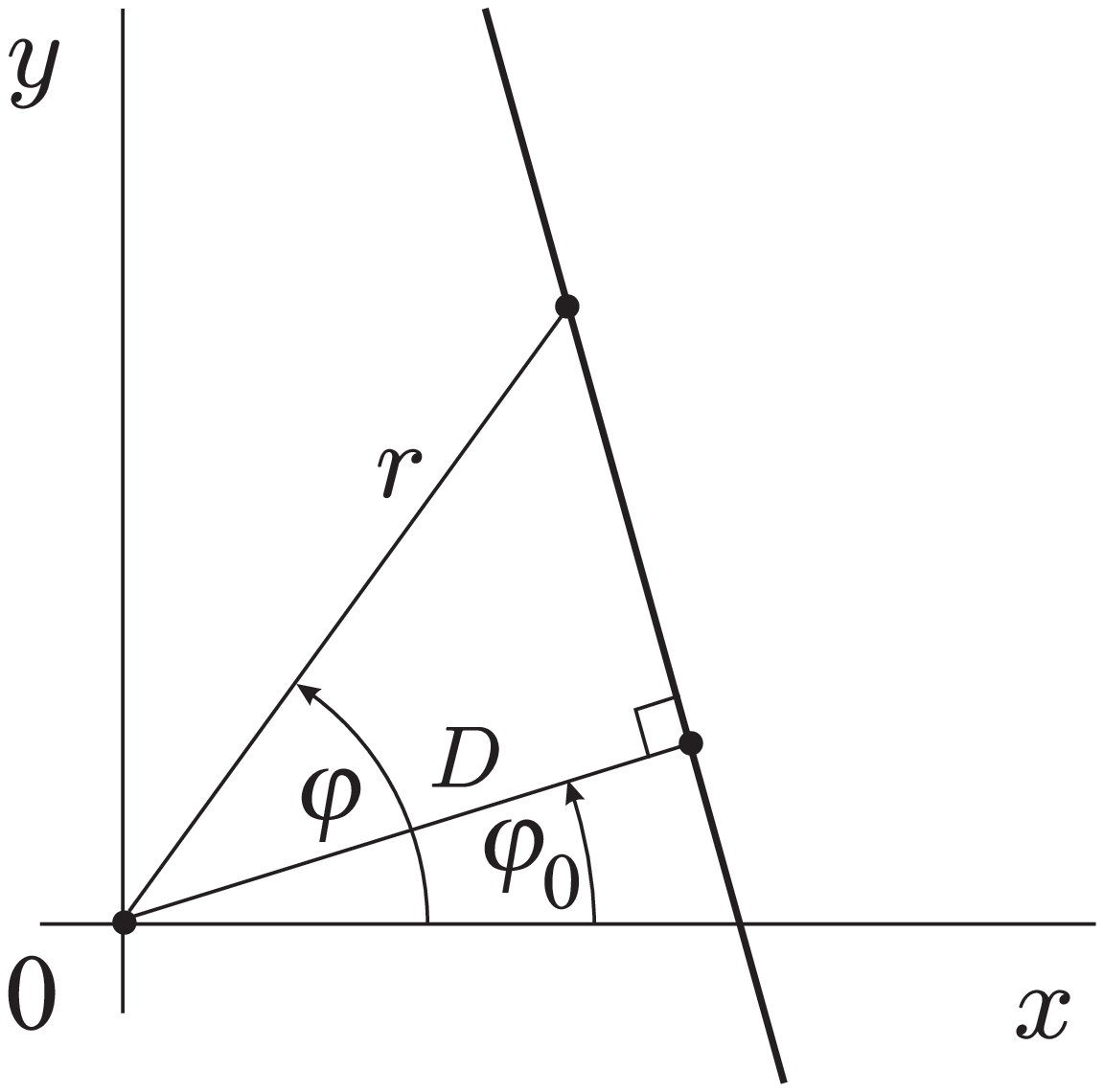,width=4cm}} 
\caption{
Rekto (\ref{reta}); ^gi estas projekcio, en ebeno $z=$ konst, de arko de cirkla helico kun nefinia radiuso.
\newline 
Figura~\ref{Reta}: The straight line (\ref{reta}); it is the projection, on a plane $z=$ const, of an arc of circular helix with infinite radius.}
\label{Reta} 
\end{figure}

\newpage%\vspace{5mm}
\ppln{{\bf {\Large 5 \hspace{5mm}Reveno al estinto}}}
\pprn{{\bf {\Large 5 \hspace{5mm}Time travel}}}
\ppln{} \pprn{}
%\ppsection[0.6ex]{Reveno al estinto\label{akcelataj3}}{Volta ao passado}                  Sekcio 5
\ppln{Ni montris ke fermitaj trajektorioj de geodezioj de (\ref{SR1}) estas cirkloj, kaj montris ke la direkto de movado estas mala al $\Omega$\,. Ni montris anka\u u ke neniu geodezio de tempa kaj nula tipo malobeas kauzecon. Nun ni montros ke (\ref{SR1}) permesas voja^ganton en ne-geodezia movado, neobeante kauzecon. Pro matematika simpleco, ni studas nur cirklajn movadojn en ebeno $z=$ konst, centrigita en akso $z$ kaj rapido $V=$ konst.}
\pprn{We have shown that the closed trajectories of geodesics of (\ref{SR1}) are circles, and that the direction of motion is opposed to $\Omega$\,. We also have shown that no timelike or null geodesic violates causality. We now show that (\ref{SR1}) allows a voyager in non-geodetic motion to violate causality. For mathematical ease, we study only circular motions in a plane $z=$ const, centered in the $z$-axis and having velocity $V=$ const.}

\ppln{$\;\;\;$ Unue ni malkovras la direkton ($\Omega$ a\u u anti-$\Omega$) de movado de voja\^ganto. Por tio, sufi\^cas konsideri ekvaciojn}
\pprn{$\;\;\;$ We first find out the direction ($\Omega$ or anti-$\Omega$) of motion of the voyager. For that, it suffices consider the equations} 

\bea 
V=r|\dd\varphi/\dd T|\,, \nonumber \hspace{1cm} (\ref{v}) 
\\ 1=\dd t/\dd T-(\Omega r^2/c^2)\,\dd\varphi/\dd T\,. \hspace{8mm} {\rm (\ref{refut}b) \nonumber} 
\eea 

\ppln{Kun $r$ kaj $V$ konstantaj, tiuj ekvacioj implicas konstantecon de $\dd\varphi/\dd T$ kaj $\dd t/\dd T$\,. Rememoru ke kauza malobeo postulas ke la sumo $\Delta t$ de infinitezimoj  $\dd t$ estu malpozitiva, en kompleta turno de voja\^ganto. \^Car $\Delta t/\Delta T=\dd t/\dd T$\,, kaj \^car $\Delta T>0$\,, tial $\dd t/\dd T<0$\,. Do (\ref{refut}b) implicas $\dd\varphi/\dd T<0$\,. Do, inercia observanto fiksata en spaca teksa\jj o, najbare la trajektorio, asertas ke la direkto de movado de voja\^ganto estas anti-$\Omega$.}
\pprn{With $r$ and $V$ constant, these equations imply constancy of $\dd\varphi/\dd T$ and $\dd t/\dd T$\,. Remember that causality violation demandas that the sum $\Delta t$ of the infinitesimals $\dd t$ be negative, in a complete turn of the voyager. Since $\Delta t/\Delta T=\dd t/\dd T$\,, and since $\Delta T>0$\,, then $\dd t/\dd T<0$\,. So (\ref{refut}b) implies $\dd\varphi/\dd T<0$\,. So an inertial observer fixed in the spatial frame, near the trajectory, asserts that the direction of motion of the voyager is anti-$\Omega$.}
 
\ppln{$\;\;\;$ Nun ni malkovras la rilaton inter $r$ kaj $V$ por ke estu ka\u uza malobeo. Por tio, sufi\^cas postuli $\dd t/\dd T<0$ kaj $V=-r\dd\varphi/\dd T$ en (\ref{refut}b); ni ricevas la kondi\^con por malobeo}
\pprn{$\;\;\;$ We now find out the relation between $r$ and $V$ so that there be causality violation. To that end, it suffices postulate $\dd t/\dd T<0$ and $V=-r\dd\varphi/\dd T$ in (\ref{refut}b); we get the condition}

\bea                                                                                \label{viol}%35
Vr>c^2/\Omega\,. 
\eea

\ppln{$\;\;\;$ Ni kalkulu $\Delta t$\,, la varion de koordinata tempo en unu kompleta turno. \^Gi estas}
\pprn{$\;\;\;$ Let us calculate $\Delta t$\,, the variation of coordinate time in a complete turn. It is}

\bea                                                                              \label{Deltat}%36
\Delta t=\frac{2\pi r}{V}\left(1-\frac{\Omega rV}{c^2}\right)\,,
\eea 

\ppln{ke klare estas negativa se ka\u uzeco estas malobeata, vidu (\ref{viol}).}
\pprn{which clearly is negative if causality is violated, see (\ref{viol}).}
 
\ppln{$\;\;\;$ Ni kalkulas anka\u u la propra intertempo $\Delta\tau$ de voja\^ganto, en unu kompleta turno. \^Car $\Delta\tau=\Delta T\sqrt{1-V^2/c^2}$\,, kaj \^car $\Delta T=2\pi r/V$\,, tial}
\pprn{$\;\;\;$ We calculate also the proper intertime $\Delta\tau$ of the voyager, in a complete turn. Since $\Delta\tau=\Delta T\sqrt{1-V^2/c^2}$\,, and since $\Delta T=2\pi r/V$\,, then}

\bea                                                                                            %37
\Delta\tau=\frac{2\pi r}{V}\sqrt{1-V^2/c^2}\,.
\eea

\ppln{Tiu rezulto koincidas kun tio de speciala relativeco.}
\pprn{This result coincides with that of special relativity.}

\vspace{5mm}
\ppln{{\bf {\Large 6 \hspace{5mm}Konkludo}}}
\pprn{{\bf {\Large 6 \hspace{5mm}Conclusion}}}
\ppln{} \pprn{}
%\ppsection[0.6ex]{Konkludo\label{konkludoFilipe}}{Conclus\~ao}                            Sekcio 6
\ppln{\^Car la sistemo de Som-Raychaudhuri estas homogena, izometra transformo forkondukas geodezion, de iu regiono de spacotempo al iu ajn regiono~\cite{PaivaMestrado}. Speciale, rezultoj pri $\dd L$, $\dd T$ kaj $V$ estas nevariantaj per tiu tranformo. Do, per simpla kalkulo, ni konstatis ke \^ciu geodezia trajektorio estas cirkla helico kun konstanta pa\^so, kaj akso paralela al akso $z$\,. Ni konstatis anka\u u ke la rapido $V$ de movado estas konstanta, je direkto mala al $\Omega$\,. La radiuso $\rho$ de helico, la rapido $V$, kaj la angulo $\alpha$ de klino de geodezia trajektorio rilatas per (\ref{novas3rho}), t.e. $\rho=(V/2\Omega)\cos\alpha$; tiu rilato klare postulas ekzisti maksimuma radiuso por geodezioj de tempa tipo.}
\pprn{Since the Som-Raychaudhuri system is homogeneous, an isometry transformation brings a geodesic from some region of spacetime to any other region~\cite{PaivaMestrado}. In particular, results about $\dd L$, $\dd T$ and $V$ are invariant under that transformation. So, via a simple calculus, we saw that every geodetic trajectory is a circular helix with fixed pitch, and axis parallel to the $z$-axis . We also saw that the velocity $V$ of the motion is constant, in the direction opposite to $\Omega$. The radius $\rho$ of the helix, the velocity $V$ and the angle $\alpha$ of slope of the geodetic trajectory relate as (\ref{novas3rho}), that is $\rho=(V/2\Omega)\cos\alpha$; this relation clearly implies existence of a maximum radius for timelike geodesics.}

\ppln{$\;\;\;$ Interesaj specialaj okazoj de helico estas: 1) Rekto en iu direkto, kaj nefinia rapido; ili estas la samtempaj geodezioj; 2) Rekto paralela al akso $z$, kun finia rapido; 3) Cirklo en ebeno $z={\rm const}$; tiuokaze $\rho=V/(2\Omega)$. En ^ci tiu lasta okazo (cirklo), estas la eblecoj: 1) $\rho=0$, por geodezio de tempa tipo de restanta objekto, kiel la polvo de modelo; 2) $\rho<c/(2\Omega)$, por alia geodezio de tempa tipo; 3) $\rho=c/(2\Omega)$, por nula geodezio, tio estas, movado de lumo; 4) $\rho>c/(2\Omega)$, por geodezio de spaca tipo; 5) $\rho\rightarrow\infty$, por samtempa geodezio.}
\pprn{$\;\;\;$ Interesting special cases of helix are: 1) Straight line in any direction, and infinite velocity; these are geodesics of simultaneities; 2) Straight line parallel to the $z$-axis, with finite velocity; 3) Cirkle in plane $z={\rm const}$, with radius $\rho=V/(2\Omega)$. In this last case (cirkle) there are the possibilities: 1) $\rho=0$, for timelike geodesics of a body at rest, such as the dust of the model; 2) $\rho<c/(2\Omega)$, for other timelike geodesics; 3) $\rho=c/(2\Omega)$, for null geodesics, that is, motion of light; 4)  $\rho>c/(2\Omega)$, for spacelike geodesics; 5) $\rho\rightarrow\infty$, for geodesics of simultaneities.}

\ppln{$\;\;\;$ Ni montris ke geodezioj de (\ref{SR1}) obeas ka\u uzecon. Sed estas bonkonata\cite{PhysLett} ke estas movadoj de akcelata materio kiuj neobeas. Ser^cante tial movadojn, ni studis cirklon en ebeno $z={\rm konst}$. Ni montris ke nur movadoj kun direkto anti-$\Omega$ kaj $\rho > c^2/(V\Omega)$ malobeas ka\u uzecon. ^Ci tiu minimuma radiuso estas la duoblo de maksimuma radiuso, $c/(2\Omega)$\,, por geodezio de tempa tipo.}
\pprn{$\;\;\;$ We have shown that geodesics of (\ref{SR1}) do not violate causality. However, it is well known~\cite{PhysLett} that there are motions of accelerated matter that violate. Looking for such motions, we studied a circle in a plane $z={\rm const}$. We have shown that only motions with anti-$\Omega$ direction and radius $\rho > c^2/(V\Omega)$ violate causality. This minimum radius is the double of the maximum radius, $c/(2\Omega)$\,, permissible for timelike geodesics.}

\ppln{$\;\;\;$ Estas tre diskutita, akcepti modelon de universo permesanta voja^ganto reveni al sia estinto. Meze de konsekvencoj de ka\u uza malobeo, estas la perdo de libera volo. Estas anka\u u malaj konkludoj, kiel tio rilata al direkto de movado. Fakte, voja^ganto en {\it tourn\'ee} de reveno al estinto perceptas ke sia angula pozicio $\varphi$ plieti^gas la\u u sia propratempo $\tau$ pligrandi^gas ($\dd\varphi/\dd\tau<0$). Do, li asertas ke la direkto de sia movado estas anti-$\Omega$. Anka\u u inercia observanto fiksata en spaca teksa^jo, en punkto najbara al trajektorio, asertas ke la direkto de movado de voja^ganto estas anti-$\Omega$, poste vidi ke $\varphi$ de voja^ganto plieti^gas la\u u la propratempo $\tau_{obs}$ de observanto pligrandi^gas ($\dd\varphi/\dd\tau_{obs}<0$)\,.}
\pprn{$\;\;\;$ It is very disputable, to accept a model universe that permits a voyager return to his past. Among the consequences of causality violation, is the loss of freedom of choice. There are also contradictory conclusions, such as the one related to the direction of motion. Indeed, a voyager in a {\it tourn\'ee} of return to the past sees his angular position $\varphi$ diminishing while his propertime $\tau$ increases ($\dd\varphi/\dd\tau<0$). So, he asserts that his direction of motion is anti-$\Omega$. Also an inertial observer fixed in the spatial frame, in a point near the trajectory, asserts that the direction of motion of the voyager is anti-$\Omega$, after seeing the position $\varphi$ of the voyager diminishing while his propertime $\tau_{obs}$ increases ($\dd\varphi/\dd\tau_{obs}<0$)\,.}

\ppln{$\;\;\;$ Nun konsideru observanton restanta en centro de la cirklo, en $r=0$. Li ne povas vidi rekte la movadon de voja^ganto, ^car la radiuso de la cirklo estas pligranda ol la diametro de geodeziaj trajektorioj de lumo, ke irus de voja^ganto al observanto. Sed ni farigas ke imagoj de la movado estu sendataj al la centra observanto per optikaj fibroj radiuse etenditaj. Pro simetrio, la tempo de aliro de ^ciu imago de voja^ganto al observanto estas la sama. ^Car voja^ganto en anti-$\Omega$ movado konstatas valorojn de $t$ pli kaj pli etaj, kaj ^car la koordinathorlo^goj estas sinkronaj al la centra horlo^go per radiusaj rektoj, tial la centra observanto vidas voja^ganto aliri kun $\dd\varphi/\dd\tau_{cen}>0$, tio estas, en direkto $\Omega$, la\u u la propratempo $\tau_{cen}$ de centra observanto pligrandi^gas.}
\pprn{$\;\;\;$ Now imagine an observer at rest in the center of the cirkle, in $r=0$. He does not see directly the motion of the voyager, because the radius of the cirkle is greater than the diameter of the geodetic trajectories of light, that would go from the voyager to the observer. But we arrange that images of the motion be sent to the central observer using optical fibers, radially extended. By symmetry, the travelling time of each image from the voyager to the observer is the same. Since the voyager in anti-$\Omega$ motion sees values of coordinate time $t$ decreasing more and more, and since these coordinate clocks are synchronous with the central clock by the radial paths, then the central observer sees the voyager moving with $\dd\varphi/\dd\tau_{cen}>0$, that is, in the $\Omega$ direction, while the propertime $\tau_{cen}$ of the central observer increases.}

\ppln{$\;\;\;$ Tamen, se horlo^go de voja^ganto estas filmata, la observanto vidas ke ^gi markas malanta\u uen la tempon, kaj vidas voja^ganton mar^sante malanta\u uen, kaj a\u udas lin paroli malanta\u uen. Tiuj strangaj perceptoj avertas la centran observanton, ke $\dd\varphi/\dd\tau_{obs}>0$ ne estas fidinda por aserti ke direkto de movado de voja^ganto estas $\Omega$.}
\pprn{$\;\;\;$ However, if the clock of the voyager is filmed, the central observer sees it counting the time backward, and sees the voyager walking backwards, and listens he speaking backward. These strange perceptions warn the central observer that $\dd\varphi/\dd\tau_{obs}>0$ is not trustful to assert that the direction of voyager's motion is $\Omega$.}

\ppln{$\;\;\;$ Tiu paradoksa fakto estas klarigita se ni analizas la sinkronon de koordinathorlo^goj. En metriko (\ref{SR1}), du koordinatholo^goj fiksitaj en la sama duonebeno $\varphi={\rm konst}$ estas sinkronaj per iu vojo en la duonebeno, kiel (\ref{refut}b) kun $\dd\varphi=0$ montras. Speciale, koordinathorlo^goj fiksitaj la\u ulonge la cirkla trajektorio de voja^ganto estas ^ciam sinkronaj al koordinathorlo^go de centra observanto en $r=0$, per radiusaj vojoj. Tio ne implicas ke tiuj horlo^goj estas sinkronaj inter ili, per vojoj en la cirklo. Vere, ili ne estas. Konsideru du najbaraj horlo^goj; en sistemo kun $\Omega>0$, la horlo^go kun plieta $\varphi$ estas malfrua rilate la alia. Fakte, postulante samtempajn momentojn ($\dd T=0$) al la horlo^goj, (\ref{refut}b) montras ke la horlo^go en plieta $\varphi$ montras valoron $(\Omega r^2/c^2)\dd\varphi$ plieta ol de alia horlo^go. Do en unu kompleta turno en anti-$\Omega$ direkto, la akumulita malfruo estos $\Delta't:=2\pi\Omega r^2/c^2$\,.}
\pprn{$\;\;\;$ This paradoxal fact is clarified if we analyze the synchronism of the coordinate clocks. In metric (\ref{SR1}), two coordinate clocks fixed in a same half plane $\varphi={\rm const}$ are synchronous by any path in the half plane, as (\ref{refut}b) with $\dd\varphi=0$ shows. In particular, the coordinate clocks fixed along the circular trajectory of the voyager are synchronous with the coordinate clock of the central observer in $r=0$, by radial paths. This does not imply that these clocks are synchronous among themselves. In fact, they are not. Take two neighbor clocks; in the system with $\Omega>0$, the clock with smaller $\varphi$ is in retard relative to the other. Indeed, if we postulate simultaneous moments ($\dd T=0$) to the clocks, (\ref{refut}b) shows that the clock with smaller $\varphi$ is counting a time value $(\Omega r^2/c^2)\dd\varphi$ shorter than the other clock. So in a complete turn in the anti-$\Omega$ direction, the accumulated retard will be $\Delta't:=2\pi\Omega r^2/c^2$\,.}

\ppln{$\;\;\;$ Nun, pensu pri voja^ganto kun tre malrapida cirkla movado ($V<<c$), en anti-$\Omega$ direkto. Poste unu turno, li maljuni^gas $\Delta\tau\approx\Delta T=2\pi r/V$\,. Sed loka tieulo maljuni^gas $\Delta t=\Delta T-\Delta't$\,, inter la du pasadoj de voja^ganto. Tiu kvanto estas ordinare pozitiva, indikante maljuni^go. Se tamen la rapido de voja^ganto estas multe pli granda, tial $\Delta T$ estas multe plieta, kaj la malfruo $\Delta't$ estas la sama. Okazos $\Delta T<\Delta't$ se rapido de voja^ganto obeas $Vr>c^2/\Omega$\,. Tiuokaze $\Delta t<0$, indikante ke lokaj tieuloj juni^gis inter la du pasadoj.}
\pprn{$\;\;\;$ Now, think about a voyager with very slow ($V<<c$) circular motion, in the anti-$\Omega$ direction. When he completes one turn, he will be $\Delta\tau\approx\Delta T=2\pi r/V$ older. But a local inhabitant will live $\Delta t=\Delta T-\Delta't$\,, between the two passages. This quantity is usually positive, indicating aging. If however the velocity of the voyager is much greater, then $\Delta T$ is much smaller, while the delay $\Delta't$ is the same. It will occur $\Delta T<\Delta't$ if the velocity of the voyager satisfies $Vr>c^2/\Omega$\,. In this case it occurs $\Delta t<0$, indicating rejuvenation of the local inhabitants between the two passages.}

%\ppparallel{\vspace{2em} % eu pus %
%\ppl{\section*{~}\vspace{-1em}} \nopagebreak % eu pus % 

%\ppr{\section*{Refer\^encias}\vspace{-1em}} \ppp \nopagebreak
%\vspace{-1.9em}} % eu pus %
\selectlanguage{esperanto}

%\ppparallel{\end{Parallel}}

\end{Parallel}

\end{document}